# Exploiting the Unique Expression for Improved Sentiment Analysis in Software Engineering Text

Kexin Sun
*State Key Lab for Novel Software Technology*
*Nanjing University*
Nanjing, China
mf20320130@smail.nju.edu.cn

Hui Gao
*State Key Lab for Novel Software Technology*
*Nanjing University*
Nanjing, China
ghalexcs@gmail.com

Hongyu Kuang*
*State Key Lab for Novel Software Technology*
*Nanjing University*
Nanjing, China
khy@nju.edu.cn

Xiaoxing Ma
*State Key Lab for Novel Software Technology*
*Nanjing University*
Nanjing, China
xxm@nju.edu.cn

Guoping Rong
*State Key Lab for Novel Software Technology*
*Nanjing University*
Nanjing, China
ronggp@nju.edu.cn

Dong Shao
*State Key Lab for Novel Software Technology*
*Nanjing University*
Nanjing, China
dongshao@nju.edu.cn

He Zhang
*State Key Lab for Novel Software Technology*
*Nanjing University*
Nanjing, China
hezhang@nju.edu.cn

*Abstract*— Sentiment analysis on software engineering (SE) texts has been widely used in the SE research, such as evaluating app reviews or analyzing developers' sentiments in commit messages. To better support the use of automated sentiment analysis for SE tasks, researchers built an SE-domain-specified sentiment dictionary to further improve the accuracy of the results. Unfortunately, recent work reported that current mainstream tools for sentiment analysis still cannot provide reliable results when analyzing the sentiments in SE texts. We suggest that the reason for this situation is because the way of expressing sentiments in SE texts is largely different from the way in social network or movie comments. In this paper, we propose to improve sentiment analysis in SE texts by using sentence structures, a different perspective from building a domain dictionary. Specifically, we use sentence structures to first identify whether the author is expressing her sentiment in a given clause of an SE text, and to further adjust the calculation of sentiments which are confirmed in the clause. An empirical evaluation based on four different datasets shows that our approach can outperform two dictionary-based baseline approaches, and is more generalizable compared to a learning-based baseline approach.

*Keywords—sentence structure, sentiment analysis, software engineering, nature language processing*

## I. Introduction

Sentiment analysis is the study of the subjectivity and polarity of a manually-written text (usually identified as positive, neutral, or negative) [1]. Modern software development process relies on a large number of manual efforts and collaborations because the scale of software is significantly larger and software development has become much more iterative [2]. Thus, the key performance indicators of software development, such as its quality, productivity, creativity, etc., will be inevitably affected by its participators' sentiments due to their indivisibility of human nature [3]. Meanwhile, the intense human collaborations of current software development are largely supported by different kinds of online tools, such as forums, communities, software repositories, and issue tracking tools. These tools then record abundant manually-written texts about the development process in the domain of software engineering (SE). These SE texts provides a valuable perspective for researchers to detect the developers' satisfaction or difficulties about the project, i.e., their positive or negative sentiments. Thus, to better support software engineering (e.g., [22]) and program comprehension (e.g., [25]) tasks, a growing body of work [19-28] applies automated sentiment analysis on SE texts from different online tools such as app stores [34-35], Stack Overflow [4, 32, 36], GitHub [29-31], and JIRA [21, 22]. These analyses are also favorable in daily SE practice because unlike the traditional approaches [5, 6, 42], they do not need direct observations or interactions on the developers, thus not likely to hinder them from their assigned development tasks.

When analyzing SE texts, the majority of the discussed work uses off-the-shelf sentiment analysis tools built on texts that are irrelevant to the SE domain, such as movie comments [7], or posts from typical social network such as Myspace [11]. To improve the performance of sentiment analysis in the SE domain, researchers further customized automated tools for SE texts by either training the particularly collected and labeled SE texts [13, 36], or building a SE-specified dictionary (e.g., mark "failure" and "exception" as neutral in SE text) [12]. Unfortunately, when analyzing the sentiments on Stack Overflow discussions to help recommend code libraries to developers, Lin et al. [36] found that no current sentiment analysis tools, even including two SE-customized tools (i.e., the domain-dictionary-based tool named SentiStrength-SE [12], and the adapted learning-based tool trained on the authors' labeled dataset from Stack Overflow), can provide reliable results of developers' sentiments in the SE texts. The reported negative results not only warn researchers about the limitations of current sentiment analysis on SE texts but also require them to further discover how developers express their sentiments in the SE texts from online collaborative tools.

In this regard, we made a close observation and found that *the expression of sentiments in SE texts are more indirect and dispersed compared to the way in the texts of common social media* (referred to as *social texts* in this paper). Specifically, we first observed that the author of an SE text often has to describe the issues that she encountered or proposed in detail before or after she expresses her sentiments, due to the overall complicacy of software tasks (such as bug fixing or comprehending code and features). Therefore, instead of assuming the entire SE text (with

---

TABLE I. THE SAMPLES TO SHOW HOW SENTISTRENGTH WORKS BASED ON ITS DICATIONARIES AND RULES WITH AN OVERALL RESULT

| Sample Text | Sent. Score $\rho$ | $\eta$ | Overall result | Dictionaries or Rules in Use | Explanation |
|---|---|---|---|---|---|
| It's a good feature. | 2 | -1 | 1 | Sentimental Word | The sentimental score of the word *'good'* is 02; so the sentence is assigned a positive score 02. |
| It's a very good feature. | 3 | -1 | 1 | Booster Word, Sentimental Word | As the booster word *'very'* before the sentimental word has the effect of +1, the sentence is assigned a positive score 03. |
| It's not good feature. | 1 | -2 | -1 | Sentimental Word Negative Word | The polarity of the sentimental word is flipped due to the use of the negation word *'not'* before sentimental word. |
| It's a good feature! | 3 | -1 | 1 | Sentimental Word "!" Rule | "!" will strengthen the sentimental strength. |
| It's a gooooood feature. | 3 | -1 | 1 | Sentimental Word Letter Repetition Rule | Repeated letters that appear more than twice above the letters required for correct spelling are used to enhance the emotional intensity of 1 unit. |

one or more sentences) as sentimental, SE-specified sentiment analysis needs to ignore clauses that are not likely to express sentiments in all sentences. We then observed that due to the more complicated writing, the sentence structures become very helpful to better understand the sentiments in SE texts, e.g., to ignore subjunctive clauses or to distinguish polysemous words.

Based on the observations, we proposed a dictionary-based approach that uses sentence structures to improve sentiment analysis on SE texts. We build our approach based on the state-of-the-art dictionary-based tool (i.e., SentiStrength [11]) instead of retraining because: (1) we can integrate our heuristics into the dictionary-based tool naturally based on our observations, and explicitly test their effects; (2) more importantly, the dictionary-based approach tends to have better generalizability on different kinds of SE texts without requiring a large amount of labeled data for training, and thus we can use four different datasets to better evaluate our observations and proposed approach. In particular, our approach consists of three major steps: (1) it preprocesses and segments a given SE text into clauses; (2) it ignores the clauses that are not likely to express sentiments according to our proposed filter rules based on the sentence structures of the SE text; (3) when identifying sentiments on the possibly sentimental clauses, our approach further uses proposed adjust rules to enhance the results of dictionary-based sentiment analysis. We evaluated our approach with the antecedent observations on four datasets that are collected from three online collaborative tools for software development: Stack Overflow, app reviews, and JIRA. The evaluation showed that our approach can substantially outperform two dictionary-based baseline approaches [7, 12] and our filter-adjust rules have a strong complementary effect to the two baselines. This result also showed that our observations, which are the basis of our proposed filter-adjust rules, are valid because they can help SentiStrength, the state-of-the-art dictionary-based tool of sentiment analysis, to achieve better performance on SE texts without modifying its dictionary of sentimental words. The evaluation also showed that our approach has a better generalizability on all four datasets than a learning-based baseline approach [13] that is trained on one dataset only.

This paper aims to improve sentiment analysis for software engineering by characterizing the unique way of expression in SE texts based on sentence structures. We name our approach as **SESSION** (**S**ent**E**nce-**S**tructure-based **S**ent**I**ment analysis for s**O**ftware e**N**gineering). This paper makes two contributions: (1) we observe and find the uniqueness of sentiment expression in SE texts; (2) we improve the accuracy of dictionary-based sentiment analysis on SE texts based on our heuristics elicited from antecedent observations by using sentence structures of the SE texts. Our tool is publicly available [43].

The rest of this paper is structured as follows. Section II introduces the background of dictionary-based sentiment analysis and our observations on sentiment expression in SE texts. Section III presents our approach. Section IV introduces the experiment and research questions. Section V answers the research questions based on the experiment results. Section VI discusses possible threats. Section VII discusses related work. Section VIII makes conclusions and refers to future work.

## II. BACKGROUND AND OBSERVATIONS ON SENTIMENT EXPRESSION IN SE TEXTS

In this section, we first introduce SentiStrength [11] which is the basis of SESSION. We then discuss the differences between SE texts and social texts when they express sentiments.

### A. How SentiStrength Works

SentiStrength is a dictionary-based sentiment classifier which is developed for common texts. It contains a series of sentiment dictionaries, including the sentimental words list, the booster word list, and the negative word list. These lists play a vital role in the computation of sentiments. The sentimental words list gives sentiment scores to the matched words. The booster word list contains words that can strengthen or weaken affected sentiment scores. The words in the negative word list are used to flip the sentimental polarity of a word right after it. For the input text, SentiStrength will assign sentiment scores to each word according to the dictionaries and use minor rules to adjust the result. We use samples in Table I to show how SentiStrength works based on its dictionaries and rules. Variables $\rho$ and $\eta$ respectively refer to the positive and negative scores for each sentence, where $+1 \leqslant \rho \leqslant +5$ and $-5 \leqslant \eta \leqslant -1$. To better detect sentiment, the default result of SentiStrength contains both two scores. Only the score of (1, -1) indicates neutrality for a text. However, it also provides a "trinary" option to output an overall sentiment that is either positive, neutral, or negative. It is worth mentioning that SentiStrength determines the sentiment scores based on the sentimental words assigned by the highest positive and negative sentiments without considering the number of clauses in the input text. This setting helps SentiStrength to focus on the most sentimental part of the input text, especially when the text size is large. We follow the same setting in our approach, but use the clauses segmented from the input text as the basis of our proposed filter-adjust rules.

## B. Different Expressions between SE Texts and Social Texts

Making close observations on SE texts and social texts, we find visible differences between two types of texts in expressing sentiments. The samples for social texts we selected are 1041 MySpace comments from the SentiStrength benchmark [11]. The samples for SE texts we selected are 4423 Stack Overflow posts from the Senti4SD benchmark [13]. Next, we will introduce our observed differences in detail.

We first find that SE texts tend to express fewer sentiments by comparing the percentage of sentimental texts from two sets of samples. For the 1041 MySpace comments, there are 938 texts manually labeled as sentimental (positive or negative). The percentage of sentimental texts is 90.1%. For the 4423 Stack Overflow posts, there are 2729 texts manually labeled as sentimental. The percentage of sentimental texts is 61.7%. In addition to the fewer sentiments, when it comes to expressing emotions, SE texts are more indirect and dispersed. We use sentimental density to reflect this characteristic of SE texts. The sentimental density $\rho$ of a text equals the number of sentimental words (according to the sentimental words list of SentiStrength) in the text $n_s$ divided by the total number of words in the text $n_w$. The average $\rho$ of the 938 MySpace sentimental texts is 0.148, while the average $\rho$ of 2729 Stack Overflow sentimental texts is 0.092. To more intuitively depict the differences, we show two samples with their $\rho$ values close to the average from the two sets of texts, respectively. The text representing MySpace is "*Thanks for the add Jeremy!! Gotta love those Macross toy pics. Sadly I don't have them anymore...* ", while the one representing Stack Overflow is "*The error occurs because of looking in the wrong environment (i.e., not inside the data frame). You could explicitly specify the but that would be ugly, awful code. Much better to use as Iselzer suggests.*" It can be observed that social texts directly express sentiments, while SE texts usually have to describe the issues first and then express the author's sentiments about the issues. An additional observation is that "error", a typical negative word for social texts, is neutral in the SE text to discuss a code issue.

We then observed that the structure of SE texts is more complicated due to the use of long and complicated sentences in SE texts to describe development-related issues. We thus measure the average length of texts in the two sets of texts by counting the number of characters. The average length of MySpace comments is 102, while the average length of Stack Overflow posts is 169. To show this difference, we also choose two texts with their length close to the average length from each of the two datasets. The text representing MySpace is "*HAPPY BIRTHDAY BEAUTIFIL... HOPE YOU SEE MANY MORE.. BETTER YET I KNOW YOU WILL...GOD BLESS YOU..STAY UP*". The whole text basically uses imperative sentences to express blessing. While the text representing Stack Overflow is "*I generally do it before importing anything. If you're worried that your module names might conflict with the Python stdlib names, then change your module names!*". The structure of this text, which contains a subjunctive clause, is more complicated.

Thus, we argue that these observed differences lead to the unreliable results provided by off-the-shelf sentiment analysis tools built on social texts, and greatly raise the difficulty to customize these tools for SE texts. The dispersed expression of sentiments requires SE-specified tools to identify whether the author is expressing sentiments in different parts of an SE text. Hence, the complicated sentence structures in SE texts become very important for us to set up filter rules to ignore possible neutral clauses, and adjust rules to enhance the output result. Our approach is built on SentiStrength with our proposed rules. The evaluation shows that our filter-adjust rules are able to customize SentiStrength for SE texts, even without updating its sentiment dictionary. For example, our approach will ignore the sentence containing the word "error" in the discussed SE-text sample instead of modifying it as "neutral" in the dictionary.

## III. PROPOSED APPROACH

We propose a three-step approach. First, we preprocess the input SE text and use Stanford CoreNLP[37] for segmentation (Step 1). Second, we use filter rules to identify whether a sentence can trigger the follow-on analysis (Step 2). Third, we use adjust rules to enhance the original output of SentiStrength (Step 3). It is worth-while noticing that our approach makes no change to SentiStrength's dictionaries. Each step will be explained with more details in the following subsections.

### A. Step 1: Preprocessing and Segmenting SE Texts

First, we adapted the preprocessing methods used by the customized tool SentiStrength-SE [12] to filter out technical words based on regular expressions and filter names containing characters such as "Dear", "Hi", "@". One difference is that we don't filter out the words fully composed by capital letters. These words are likely to express an exaggerated sentiment, rather than be just part of technical texts. We also keep exclamation marks as part of the input for Step 3. The text "*FEAR!!!!!!!!!!!* " is a good sample to illustrate the above two differences. Besides, we will also filter out the words surrounded by the following brackets "[]", "{}", "<%%>", and double quotation marks because we think that these words are more likely to be quotations, examples, or technical words and to not express sentiments. For example, in the sentence "*CREATE TABLE [[With Spiteful]]...*", "spiteful" is a negative word but it is part of the table's name and doesn't express sentiments. Similarly, the negative word "tommyrot" in the sentence "*It is actually spelled "tommyrot".*" does not indicate negative sentiment because it is quoted as an example. Additionally, the sentence with underline symbols, e.g., "CODE_FRAGMENT", will be filtered too because this symbol is also a feature of technical text.

Second, to deal with SE texts which have more complicated sentence structures, we introduce Stanford NLP to segment, instead of following SentiStrength to segment texts according to punctuation marks only. Our segmentation first divides the whole text (named as *paragraph*) into multiple *sentence*s. It then divides each sentence into clauses based on punctuations and conjunctions such as "because", "but", and "so". Furthermore, we use Stanford POS tagger to annotate each *word* in the clauses of each sentence with its *part of speech(POS)* tagging. The preprocessed, segmented, and tagged SE texts lays the foundation of the following steps of our approach.

### B. Step 2: Matching Patterns to Trigger Follow-on Analysis

To distinguish whether the author is expressing sentiments or describing issues, we propose our filter rules. Specifically,

any sentence that did not fit the following three patterns will be filtered out. Only the sentence that matches at least one defined pattern will be considered as likely to express sentiments, and will go to the next step for calculating its sentiment scores. A detailed description of patterns is as follows.

*1) Direct Sentiment Pattern.* A given sentence fits Direct Sentiment Pattern when it matches just one of the following six situations: (1) it contains the exclamation marks; (2) it contains emoji recorded in the SentiStrength's emoji list, such as " **:)** "; (3) it contains interjection word according to the tagged POS, such as "wow"; (4) it contains the four four-letter curse words that respectively start with letters "fu", "da", "sh", and "he"; (5) at least one of its given clauses starts with a sentimental word (except "please" and "plz"); (6) it is an imperative sentence and has a sentimental density larger than 0.3.

Intuitively, the first four situations indicate that the authors strongly expressed their sentiments. Meanwhile, we propose the fifth and the sixth situations to deal with imperative sentences. The fifth situation is proposed to cover the following two sample sentences: "*Thanks for your patience.*" and "*Owen, thanks for the slides.*". We exclude "please" and "plz" in the fifth situation because they are more likely to express requests instead of their intended positive sentiments. The sixth situation is proposed to cover the following sample sentence: "*Sounds good.*" . How to calculate the sentimental density for each sentence is discussed in Section II.B.

*2) Decorated Sentiment Pattern.* A given sentence fits Decorated Sentiment Pattern when it contains a sentimental word that is an adverb, or it contains a sentimental word that is decorated by an adverb (implying that this sentimental word must be a verb or an adjective). We suggest that when using sentimental adverbs, or adverbs to decorate a sentimental word, the author is determined to express her sentiments in the text because adverbs are used to indicate degree or scope. For example, in the sentence "*This is very frustrating.*", the adverb "very" indicates a deeper frustration (i.e., negative sentiment). While in the sentence "*The performance degrades horrendously*", the adverb "horrendously" indicates the degree of performance degradation is too large and thus showing the author's negative sentiment as well. Furthermore, for the three adverbs "always", "even", and "still", we will find decorated sentimental words from these words to the end of the sentence because they have a wider coverage based on their semantics. Finally, we treat "how", "sort of", and "enough" (after sentimental words) as adverbs because they are also highly likely to indicate the degree or scope of potential sentiments.

*3) "About Me" Pattern:* A given sentence fits "About Me" Pattern when it matches the following three situations: (1) its subject is "I" and it contains a sentimental word (e.g. "*I like…*"); (2) it contains a sentimental verb followed by the object "me" (e.g. "*…confuse me*"); (3) it contains a sentimental adjective or noun that follows "me" (e.g. "*…make me confused*"); (4) it contains a sentimental word that is decorated by "my" (e.g., "*This was my bad.*"). We propose the four situations because we suggest that the author is determined to express her sentiments in the first-person view. On contrary, the third-person view is usually more likely to describe a fact, instead of expressing sentiments. For example, the sentence "*he hates p tags, clearly*" is manually labeled as neutral.

*4) "Judgement" Pattern:* A given sentence fits "Judgement" Pattern when it contains the following four sentence structures (1) "be verb + sentimental adjectives/nouns" (e.g., "*It's ugly and inefficient*"); (2) "pronoun + sentimental verb" (e.g., "*This sucks so much.*"); (3) "get + sentimental word" (e.g., "*The problem just gets worse.*"); (4) "sentimental nouns + be verb" (e.g., "*The biggest reason for failure is your carelessness*"); (4) "a/an/the + adjective + noun" ("*It has an excellent command line interface.*"). We argue that the author usually expresses her sentiments when she makes a judgement to other things or people, and the five proposed situations can largely cover the potential judge-and-express scenarios.

C. Step 3: Adjusting the Sentiment Analysis

We argue that sentence structures are also helpful to better understand expressed sentiments in SE texts. So we propose to adjust rules based on SentiStrength to further enhance the results.

*1) Recognizing Subjunctive Mood:* Subjunctive mood expresses the author's subjective wishes, suspicions, suggestions, or hypotheses, but does not express real sentiments. Therefore, we ignore the sentimental words occurred in clauses of subjunctive mood. Our approach identifies subjunctive mood by recognizing "if" and "unless" as conditional adverbials in the clauses of a given sentence. We will not identify the sentiments in these clauses. For example, in the sentence "*If you're really worried about this, Java is not the language for you.*" the negative sentimental word "worried" is in the subjunctive clause, so it reflects no facts and does not express the author's sentiments.

*2) Identifying Polysemous Words by the Sentence Structure:* SentiStrength assigns a sentimental score to each sentimental word. However, when sentimental words express different meanings according to the different sentence structures, a single sentimental score will lead to possibly wrong results. During our observations, we summarized several polysemous words that can easily lead to mistakes. These words are categorized into two groups. We then confirm the meaning of first-group words based on the POS tags, and the meaning of second-group words based on their collocations with other words.

The first group of polysemous words that can be confirmed by the POS tags is as follows:

**Like:** SentiStrength detects this word as positive. In the sentence "*I like playing with you*", the word "like" is positive and it means that the subject prefers to do something. However, in the sentence " *it looks like this.* ", its meaning is close to "similar to" and it doesn't express positive sentiments. When "like" means " similar to", its POS is a preposition. So when its POS is preposition, we do not mark this word as positive, but as neutral instead.

**Pretty and Super:** SentiStrength detects these words as positive. In the sentence "*She is pretty.* ", the word "pretty" is positive and it means someone is attractive. However, in the sentence " *I'm pretty sure* " its meaning is close to "very" and it doesn't express positive sentiments. When "pretty" means "very", its POS is an adverb. So when its POS is an adverb, we do not mark this word as positive but as neutral. It will also play the role of booster words that can strengthen the intensity of the following sentiment word, like "very". "Super" is similar to "pretty". When its POS is an adverb and it is used to indicate something with a high or extreme degree, we detect it as neutral and it will play the role of booster words as well.

**Block and Force:** SentiStrength detects these words as negative. In sentences " *Lack of training acts as a block to progress in a career.*", the word "block" is negative and it means something that makes movement or progress difficult or impossible, but in sentences similar to " *I'm sure at first the code blocks*", it means a quantity of something that is considered as a single unit and does not express any negative sentiments. When "block" means " a unit", its POS is a noun. So when its POS is noun, we do not mark it as negative but as neutral. "Force" is similar to "block". When its POS is a noun, it means physical strength and we mark it as neutral instead of negative.

The second group of polysemous words that can be confirmed by their collocations with other words is as follows:

**Lying:** SentiStrength detects the word as negative. In the sentence "*He was lying.*", the word "lying" is negative and it means something deviating from the truth, but in sentences similar to "*It's lying all over the internet.*", its meaning is close to "be in" and it does not express negative sentiments. When "lying" means "be in", it is often used with prepositions, except "to" (excluding the phrase "lie to"). So when we recognize this collocation, we do not mark it as negative but as neutral.

**Spite and Kind:** SentiStrength detects the word "spite" as negative, but in the phrase "in spite of", the whole phrase represents a turning relationship and expresses no negative sentiments. So when found in this phrase, we do not mark it as negative but as neutral. "Kind" is similar to "spite". In the phrase "kind of", the meaning of the phrase is close to " to some extent" and the phrase expresses no positive sentiments. So we do not detect it as positive but as neutral when found in this phrase.

**Miss:** The word "miss" is assigned both a positive score 02 and a negative score 02 by SentiStrength because when its meaning is close to "remember fondly", it is frequently used to express sadness and loves simultaneously. However, when its meaning is close to "notice something not there", it expresses negative sentiments in SE texts. According to our observation, when it means "remember fondly", it is often followed by personal pronouns. When it means "notice something not there", it is followed by the object. Therefore, we will check the object of this word, only when its object is a personal pronoun, we will calculate its positive and negative sentiments at the same time.

*3) Dealing with Negations.* The original rule about negations in SentiStrength will flip the polarity of a sentimental word by multiplying a factor of -0.5 when a negation word is right in front of it. This rule overcompensates and ignores too many negation scenarios, especially for SE texts. For example, the sentiment of this text "*not to worry, it was a permissions issue with the file.*" will be identified as positive according to the original negation rule, but it is labeled as neutral. Instead in our approach, the words in the negation words list and the words ending with "'t" (e.g., "isn't") will neutralize the sentiment of the words within the following three words ("to" excluded). We also add three more words "nothing", "no", and "without" (not in the original negation list of SentiStrength) to neutralize the sentiment of the first word ("to" excluded) right behind them. The limited negation scope of the added three negation words is because their POS are nouns or prepositions, while the negation words in the original list or ending with "'t" are auxiliary verbs.

TABLE II. THE ANALYSIS (WITH TRINARY OUTPUT) OF SENTISTRENGTH

| Sentence | Senti. Score | |
| --- | --- | --- |
| | ρ | η |
| This app is *really good*[2] [+1 booster word] in *spite* [-4] of some (minor) *shortcomings*[-2] . | 3 | -4 |
| Its font sizes will get bigger or smaller to fit the space for them and i *don't like*[2] [*-0.5 approx. negated multiplier] . | 2 | -1 |
| If the *problem*[-2] solved, I think it will be more practical . | 1 | -2 |
| Overall ,it's a *good*[2] app though . | 2 | -1 |
| **Overall result = -1 as Max(ρ) < Max(abs(η))** | | |

TABLE III. THE ANALYSIS (WITH TRINARY output) OF SESSION

| Sentence | Senti. Score | |
| --- | --- | --- |
| | ρ | η |
| [fit "Decorated sentiment Pattern"] This app is *really good*[2] [+1 booster word] in *spite* [polysemous words] of some (minor) *shortcomings*[-2] . | 3 | -2 |
| [fit "'About Me' Pattern"] Its font sizes will get bigger or smaller to fit the space for them and i *don't like* [neutralized by negations] . | 1 | -1 |
| [does not fit any pattern] If the problem solved, I think it will be more practical . | 1 | -1 |
| [fit "'Judgement' Pattern"] Overall ,it's a *good*[2] app though . | 2 | -1 |
| **Overall result = 1 as Max(ρ) > Max(abs(η))** | | |

*D. Summary through a Sample SE Text*

We now use the following sample SE text to show how SESSION works: " *This app is a really good in spite of some (minor) shortcomings. Its font sizes will get bigger or smaller to fit in the space allocated for them which I don't like. If you can solve the problem, I believe it will be more practical. Overall, it's a good app though.*". The sentiment of this text is manually labeled as positive. The analysis and results from original SentiStrength are shown in Table II, while the analysis and results from SESSION is shown in Table III. It can be observed that, based on our proposed filter rules and adjust rules (Step 2 and Step 3) that rely on the segmentation and POS tagging of preprocessed SE texts in Step 1, SESSION correctly identifies the positive sentiment for this text, while SentiStrength is misled by the text to wrongly identify its sentiment as negative.

IV. EXPERIMENTAL SETUP

We now introduce our experimental setup to evaluate our approach. Section IV.A introduces the four datasets of SE texts for the evaluation. Section IV.B defines metrics for evaluating the performance of the proposed approach. Section IV.C introduces our research questions and the design of experiments.

*A. The Benchmark with Four Datasets*

We first bring in the benchmark that Lin et al. studied and reported that no current sentiment analysis tools can provide reliable results of sentiments expressed in the SE texts [36]. It consists of three datasets that are built on 1500 Stack Overflow discussions, 341 app reviews, and 926 JIRA comments, respectively. We then introduce the fourth dataset that is built on 4423 Stack Overflow posts by Calefato et al. to propose a learning-based approach of sentiment analysis on SE texts. Table IV reports the total number of texts, and the number of positive, neutral, and negative texts for each dataset.

TABLE V.  THE PERFORMANCE OF SESSION AND THREE BASELINES ON THE FOUR DATASETS

| Dataset | Tool | overall accuracy | positive P | positive R | positive F | neutral P | neutral R | neutral F | negative P | negative R | negative F |
|---|---|---|---|---|---|---|---|---|---|---|---|
| Stack Overflow 4423 | SentiStrength | 81.55% | 88.90% | 92.34% | 0.906 | 92.76% | 63.58% | 0.754 | 66.83% | 93.18% | 0.778 |
| | SESSION | 86.30% | 90.15% | 94.70% | 0.924 | 90.19% | 75.97% | 0.825 | 77.87% | 90.18% | 0.836 |
| | SentiStrength-SE | 78.86% | 90.47% | 82.06% | 0.861 | 72.74% | 77.80% | 0.752 | 74.80% | 76.29% | 0.755 |
| | Senti4SD | **95.27%** | 97.25% | 97.45% | 0.974 | 95.02% | 93.51% | 0.943 | 93.15% | 95.01% | 0.941 |
| Stack Overflow 1500 | SentiStrength | 68.00% | 19.28% | 36.64% | 0.253 | 86.20% | 74.98% | 0.802 | 36.74% | 44.38% | 0.402 |
| | SESSION | **78.13%** | 30.89% | 29.01% | 0.299 | 85.10% | 89.67% | 0.873 | 54.10% | 37.08% | 0.44 |
| | SentiStrength-SE | 78.00% | 31.18% | 22.14% | 0.259 | 82.72% | 92.86% | 0.875 | 50.00% | 19.66% | 0.282 |
| | Senti4SD | 76.93% | 27.59% | 30.53% | 0.29 | 83.11% | 90.51% | 0.867 | 62.07% | 20.22% | 0.305 |
| App Reviews | SentiStrength | 67.45% | 71.81% | 87.63% | 0.789 | 4.76% | 4.00% | 0.043 | 70.97% | 50.77% | 0.592 |
| | SESSION | **68.62%** | 76.17% | 87.63% | 0.815 | 9.76% | 16.00% | 0.121 | 77.91% | 51.54% | 0.62 |
| | SentiStrength-SE | 61.58% | 74.15% | 81.72% | 0.777 | 9.59% | 28.00% | 0.143 | 80.95% | 39.23% | 0.528 |
| | Senti4SD | 63.93% | 71.24% | 86.56% | 0.782 | 9.80% | 20.00% | 0.132 | 81.25% | 40.00% | 0.536 |
| JIRA Issue | SentiStrength | **81.21%** | 86.03% | 93.45% | 0.896 | — | — | — | 98.16% | 75.63% | 0.854 |
| | SESSION | 80.56% | 93.13% | 93.45% | 0.933 | — | — | — | 98.55% | 74.69% | 0.85 |
| | SentiStrength-SE | 77.21% | 95.26% | 90.00% | 0.926 | — | — | — | 99.34% | 71.38% | 0.831 |
| | Senti4SD | 57.88% | 81.55% | 86.90% | 0.841 | — | — | — | 99.65% | 44.65% | 0.617 |

B. Metrics

We first leverage three metrics to measure the accuracy of sentiment analysis for each of the three sentimental polarities (i.e., positivity, negativity, and neutrality). Given a set $S$ of texts, *precision (P)*, *recall (R)*, and *F-measure (F)* for a particular sentimental polarity is calculated as follows:

$$P = \frac{|S_c \cap S'_c|}{|S'_c|} \quad R = \frac{|S_c \cap S'_c|}{|S_c|} \quad F = \frac{2 \times P \times R}{P + R} \quad (1)$$

where $S_c$ represents the set of texts having sentimental polarity $c$, and $S'_c$ represents the set of texts classified to have sentimental polarity $c$ by a tool. F-measure is the weighted harmonic mean of precision and recall. A higher F-measure means both precision and recall are high, and the tool performs better. We further introduce the overall accuracy of sentimental analysis on the set $S$ for all of the three sentimental polarities with metric *Overall Accuracy* calculated as follows:

$$Overall\ Accuracy = \frac{\sum_{c \in polarities}|S_c \cap S'_c|}{|S|} \quad (2)$$

where we accumulate the numbers of texts in $S'_c$ which have the same sentimental polarity "c" in $S_c$ for all three polarities, and then calculate the proportion of it in the given set $S$ of texts.

C. Research Question

In this paper, we aim to study whether sentence structures can effectively improve the performance of sentiment analysis in SE texts. Therefore, we propose the following three research questions:

TABLE IV.  DATASETS USED FOR OUR EVALUATION

| Dataset | sentences | positive | neutral | negative |
|---|---|---|---|---|
| Stack Overflow 4423 | 4423 | 1527 | 1694 | 1202 |
| Stack Overflow1500 | 1500 | 131 | 1191 | 178 |
| App Reviews | 341 | 186 | 25 | 130 |
| JIRA issue | 926 | 290 | 0 | 636 |

*RQ1: Can our proposed approach outperform the baseline in analyzing sentiments for SE texts?*

*RQ2: How much contribution do our filter rules make?*

*RQ3: How much contribution do our adjust rules make?*

To study *RQ1*, we introduce the following three baselines: (1) *SentiStrength* [11], the state-of-the-art dictionary-based tool and the basis of our approach; (2) *SentiStrength-SE* [12], a representative dictionary-based tool that builds a new dictionary specified for SE texts; (3) *Senti4SD* [13], a representative SE-Customized, learning-based tool that is trained on the **Stack Overflow 4423** dataset (also part of our evaluated datasets). Based on the comparison with the three baseline approaches, we expect to find out whether our approach can have a better performance, as well as whether our observations about the uniqueness of sentiment expression in SE texts are valid. To study *RQ2* and *RQ3*, We will run SentiStrength with our filter rules only (*SS + Filter*) and with our adjust rules only (*SS + Adjust*) on the four database, respectively, to further compare their performances with SentiStrength and SESSION.

V. RESULTS AND DISCUSSIONS

A. RQ1: Can our proposed approach outperform the baseline in analyzing sentiments for SE texts?

Table V shows the performances of the evaluated four approaches. First, we compare the performance of SESSION with SentiStrength. We found that the overall accuracy of SESSION in **Stack Overflow 4423**, **Stack Overflow 1500**, and **App Reviews** is better than that of SentiStrength. Its overall accuracy on **Stack Overflow 1500** can be 10% higher than that of SentiStrength. Our previous observations show that social texts are more sentimental and their expression is more direct than SE texts. This difference makes SentiStrength tend to output more positive and negative results. This tendency can be observed through the low recall of identified neutral texts achieved by SentiStrength in Table V. On the other hand, we propose filter rules and adjust rules to address the issue that the sentiments expression in SE texts is more indirect and dispersed. Our approach thus achieves 12% more recall than SentiStrength on **Stack Overflow 4423**. The proposed filter-adjust rules also contribute to help our approach outperform SentiStrength in the F-Measures of all three sentiment polarities on the evaluated datasets, except **JIRA Issue**. Unlike the two datasets from Stack Overflow, **JIRA Issue** has no neutral texts. Thus, it leaves little room for our filter-adjust rules to work. However, our approach still outperforms SentiStrength in the F-Measure of positive sentiments on **JIRA Issue**, while only performs slightly worse in the F-Measure of negative sentiments. Because **JIRA Issue** contains about two times more negative texts than its positive texts and no neutral texts, SESSION thus performs slightly worse in the overall accuracy (further discussions are in the end of this section). We then use sample texts (from the four datasets) shown in Table VI to demonstrate how SESSION outperforms SentiStrength. In the table, the first sentence is identified as positive by SentiStrength because of "!", while SESSION can filter out "!=" as part of technical text. The second sentence is identified as negative by SentiStrength because of "worried", while SESSION locates its subjunctive mood and identifies it as neutral. Because of "hate", the third sentence is identified as negative by SentiStrength, while this sentence cannot fit any patterns in our filter rules and SESSION identifies it as neutral.

TABLE VI. SAMPLES FOR COMPARING SESSION (SN) WITH SENTISTRENGTH (SS, M STANDS FOR MANUAL LABEL)

| Sentence | M | SN | SS |
|---|---|---|---|
| It's pretty easy to prevent aliasing by adding a conditon *a != *b. | 0 | 0 | 1 |
| If you're really worried about this, Java is not the language for you | 0 | 0 | -1 |
| why do people hate anonymous block initializers | 0 | 0 | -1 |

Second, we compare their performances between SESSION and SentiStrength-SE. From Table V we can find that SESSION outperforms SentiStrength-SE in overall accuracy and almost in all the other metrics on the four datasets, except the recall and F-Measure of neutral sentiments on **Stack Overflow 1500** and **App Reviews**, and the recall of neutral sentiments on **Stack Overflow 4423** where SESSION slightly performs worse. Both approaches actually exploited the neutral tendency of SE texts. SentiStrength-SE chooses to establish a SE-domain-specified dictionary, while our approach chooses to use filter-adjust rules to enhance SentiStrength. We found that the overall accuracies for SentiStrength-SE and SESSION differ little on **Stack Overflow 1500**. However, to cope with the neutral tendency, the updated sentimental word list of SentiStrength-SE is shortened to 550 words only, while the original list in SentiStrength has more than 2,000 words. Consequently, SentiStrength-SE covers much fewer possible positive and negative sentiments than SESSION and SentiStrength. Sentences like "*I'm loving.*" will not be identified as sentimental by SentiStrength-SE because it lacks the sentimental word "loving" in its list. We then argue that our observations and proposed filter-adjust rules better exploit the unique expression of sentiments in SE texts. We use sample texts shown in Table VII to demonstrate how SESSION outperforms SentiStrength-SE. In the table, the first sentence is classified as neutral by SentiStrength-SE because it will delete "!" during preprocessing. The preprocessing rule of SESSION will keep "!" so the text won't be misclassified. The second sentence is classified as negative by SentiStrength-SE because the word "messagebox" matches its wildcard "mess*" which has negative score 02 in SentiStrength-SE's sentimental word list. This word doesn't match in SESSION's sentimental word list so the text won't be misclassified. The third sentence is classified as negative by SentiStrength-SE because of "afraid", while this sentence cannot fit any patterns in our filter rules and SESSION identifies it as neutral.

TABLE VII. SAMPLES FOR COMPARING SESSION (SN) WITH SENTISTRENGTH-SE (SE, M STANDS FOR MANUAL LABEL)

| Sentence | M | SN | SE |
|---|---|---|---|
| Joei get it! i guess you are right | 1 | 1 | 0 |
| How to correctly print a CString to messagebox? There is nothing appear.. | 0 | 0 | -1 |
| Are you afraid of a trademark lawsuit? | 0 | 0 | -1 |

Third, we compare their performances between SESSION and Senti4SD. From Table V, we found that Senti4SD significantly outperforms SESSION in its training set **Stack Overflow 4423**. We think this result is reasonable because, with the help of improved feature engineering to cover more implicit facts [13], Senti4SD can better predicate the sentiments in the SE texts, especially from **Stack Overflow 4423** where Senti4SD fine-tunes the parameters of its trained SVM model for classification. However, when applied to other datasets, the performance of Senti4SD begins to decrease. Its performance on **Stack Overflow 1500**, the dataset similar to its training set, is lower than SESSION. The same comparison can also be observed on **App Reviews**. Moreover, its overall accuracy on **JIRA Issue** is only 57.88% and its negative recall is only 44.65%. On the other hand, SESSION is able to achieve balanced recall and precision for all sentiments. The recall of negative text is about 10%-30% higher than that of Senti4SD. We argue that SESSION achieves a comprehensively better performance than Senti4SD, especially in the generalizability.

Our overall observation on the evaluation shows that tools (SentiStrength-SE, Senti4SD) developed from software engineering texts can often achieve higher precision in sentiment texts, but has to suffer the cost of a lower recall. Tools (SentiStrength) developed for social texts can often achieve higher recall in sentiment texts, but to suffer the loss of precision. In contrast, our tools, which exploit the unique expression of sentiments in SE texts based on sentence structures, can achieve a good and balanced performance in precision and recall, and a better generalizability when compared to a learning-based tool.

*B. RQ2: How much contribution do our filter rules make?*

The results of SS + Filter are shown in Table VIII. Comparing the data of SS + Filter with SentiStrength, we can find that in two Stack Overflow datasets, the overall accuracy of SS + Filter is better than that of the original tool. In **Stack Overflow 4423**, the overall accuracy of SS + Filter is 2.13% higher, and In **Stack Overflow 1500**, the overall accuracy of SS + Filter is 6.93% higher. These rules can effectively improve the precision of sentimental texts and the recall of neutral texts and especially better at neutral texts. The neutral F-measure of the tool with filter rules is all higher than that of the original tool and that of the tool with filter rules. In the other two datasets, the improvement of the overall accuracy that filter rules can bring is not high. Because there are few neutral texts on these two datasets, filter rules, which are better at neutral texts, are difficult to bring improve. In summary, because the F-measures of SS + Filter are almost all better than SentiStrength, we can say that filter rules can actually bring improvements. However, its improvement will be a little unstable when analyzes datasets with too many sentimental texts.

TABLE VIII. ANALYZING PERFORMANCES OF RULE-FILTER AND RULE-ADJUST RESPECTIVELY

| Dataset | tool | overall accuracy | positive | | | neutral | | | negative | | |
|---|---|---|---|---|---|---|---|---|---|---|---|
| | | | P | R | F | P | R | F | P | R | F |
| Stack Overflow 4423 | SentiStrength | 81.55% | 88.90% | 92.34% | 0.906 | 92.76% | 63.58% | 0.754 | 66.83% | 93.18% | 0.778 |
| | SS + Filter | 83.68% | 90.06% | 91.94% | 0.91 | 90.56% | 70.78% | 0.795 | 71.30% | 91.35% | 0.801 |
| | SS + Adjust | 84.08% | 89.00% | 94.89% | 0.919 | 92.06% | 68.42% | 0.785 | 72.33% | 92.43% | 0.812 |
| | SESSION | **86.30%** | 90.15% | 94.70% | 0.924 | 90.19% | 75.97% | 0.825 | 77.87% | 90.18% | 0.836 |
| Stack Overflow 1500 | SentiStrength | 68.00% | 19.28% | 36.64% | 0.253 | 86.20% | 74.98% | 0.802 | 36.74% | 44.38% | 0.402 |
| | SS + Filter | 74.93% | 23.31% | 29.01% | 0.259 | 85.12% | 85.47% | 0.853 | 48.23% | 38.20% | 0.426 |
| | SS + Adjust | 73.87% | 27.43% | 36.64% | 0.314 | 86.23% | 82.54% | 0.843 | 41.62% | 43.26% | 0.424 |
| | SESSION | **78.13%** | 30.89% | 29.01% | 0.299 | 85.10% | 89.67% | 0.873 | 54.10% | 37.08% | 0.44 |
| App Reviews | SentiStrength | 67.45% | 71.81% | 87.63% | 0.789 | 4.76% | 4.00% | 0.043 | 70.97% | 50.77% | 0.592 |
| | SS + Filter | 67.45% | 75.36% | 85.48% | 0.801 | 10.00% | 16.00% | 0.123 | 74.44% | 51.54% | 0.609 |
| | SS + Adjust | **69.21%** | 73.45% | 89.25% | 0.806 | 8.33% | 8.00% | 0.082 | 74.73% | 52.31% | 0.615 |
| | SESSION | 68.62% | 76.17% | 87.63% | 0.815 | 9.76% | 16.00% | 0.121 | 77.91% | 51.54% | 0.62 |
| JIRA Issue | SentiStrength | 81.21% | 86.03% | 93.45% | 0.896 | —— | —— | —— | 98.16% | 75.63% | 0.854 |
| | SS + Filter | 80.35% | 87.91% | 92.76% | 0.903 | —— | —— | —— | 97.94% | 74.69% | 0.847 |
| | SS + Adjust | **82.18%** | 91.28% | 93.79% | 0.925 | —— | —— | —— | 98.19% | 76.89% | 0.862 |
| | SESSION | 80.56% | 93.13% | 93.45% | 0.933 | —— | —— | —— | 98.55% | 74.69% | 0.85 |

*C. RQ3: How much contribution do our adjust rules make?*

The data of SS + Adjust is shown in Table VIII. We can find that the overall accuracy of SS + Adjust in four datasets is all better than SentiStrength. It can also effectively improve the precision of sentimental texts and the recall of neutral texts. The F-measures of SS + Adjust are all better than SentiStrength, so we can say that adjust rules can actually bring improvements. Compared to filter rules, they are better at sentimental texts. The positive F-measures and negative F-measures of the tool with adjust rules are almost all higher than the original tool and the tool with filter rules. For sentimental texts, adjust rules can improve the precision without losing too much recall. In **Stack Overflow 4423**, the positive precision (89.00%) of SS + Adjust and that (90.06%) of SS + Filter are similar. But the positive recall (94.89%) of SS + Adjust is higher than that (91.94%) of SS + Filter. In addition, we also find that the improvement brought by filter rules will be a little less when analyses datasets with too many sentimental texts. In two Stack Overflow datasets which have more neutral texts, the overall accuracies of SS + Adjust are 2.53% and 5.87% higher than the original tool respectively. In the other two datasets with more sentimental texts, the overall accuracies of SS + Adjust are 1.76% and 0.97% higher than the original tool, respectively.

In summary, the two sets of rules for our approach can both bring improvements because they can effectively improve the precision of sentimental texts and the recall of neutral texts. Because our rules are based on the observation that SE texts are more indirect and complicated than social texts, they will be more helpful when analyses datasets with more neutral texts. To be more specific, Table VIII shows that SESSION (with both filter rules and adjust rules) performs best on both **Stack Overflow 4423** and **Stack Overflow 1500**, while SS + Adjust performs best on both **App Reviews** and **JIRA Issue**. This observation shows that although our filter rules are better at handling neutral texts, they may also conduct a loss of the sentiment context when filtering out sentences that cannot match any patterns, especially compared to the adjust rules which performs more stably on all SE texts. However, when applied on **App Reviews** and **JIRA Issue**, our filter rules only decrease the overall accuracy by 0.59% and 1.62%, respectively. Because these two data sets only respectively contains 7% and 0% neutral texts, indicating that our filter rules have little room to work, we suggest that the possible loss of sentiment context caused by our filter rules are not significant. We then suggest that due to the more indirect and dispersed nature of sentiment expression in SE texts, both our filter rules and adjust rules are helpful for sentiment analysis on SE texts generated by the online tools for SE in practice, where neutral texts are likely to take a big part.

Additionally, we made three more observations on the experiment results. First, our experiment results for SentiStrength on the three datasets that Lin et al. also studied are a little different from the results in their paper [36]. We found that it is because Lin et al. uses the sign of the sum of positive and negative scores from SentiStrength to get the overall polarity, while our approach uses the in-built "trinary" option of SentiStrength to output the overall polarity. By comparison, we found that our results for SentiStrength are slightly better, and thus we make no bias when comparing with SentiStrength. Second, the improvement of our approach, though is balanced and stable on all datasets, is still not high. We think this situation is caused by our conservative choice of using sentence-structure-based rules to collaborate with SentiStrength. In future work, we plan to carry out a deeper study on how developers express their sentiments in SE texts and to carefully establish SE-specified dictionaries by consulting existing work [45]. Third, we found that the standards of manual labeled sentiments can vary in different datasets. During our research, we had a candidate polysemous word "work". When "work" is an intransitive verb, it means "to effect something" and can be viewed as a positive sentimental word. However, **Stack Overflow 4423** favors this candidate word, while **Stack Overflow 1500** tends to be the opposite, and thus we finally exclude this word from the adjust rules of SESSION. Our further investigation shows that although the two datasets are both created from Stack Overflow, the participants who label sentiments for **Stack Overflow 1500** tend to favor the neutral texts instead of either positive texts or negative texts. For example, in this dataset the typical positive texts such as "I appreciate your help", and the typical negative texts such as "I suspect why the decision is made", are both labeled as neutral. A possible explanation is that, in the participants' opinions, the sentiments of these texts, either identified as positive or negative, are not convincing enough to indicate the real status of the potentially related SE tasks to other developers. The similar situation also occurs in **App Reviews** where 25 texts are manually labeled as neutral, while they contain considerable number of sentimental words. These results of our investigation is able to explain why SESSION and all baseline approaches do not perform well on the positive and negative texts of **Stack Overflow 1500**, and the neutral texts of **App Reviews**. We thus suggest that it would be favorable if the SE community could agree on a unified standard for manually labeling sentiments on SE texts to help researchers (including us) establish more consistent datasets that aim to enhance the research of sentiment analysis in the SE domain.

## VI. THREATS TO VALIDITY

**Internal Threats.** A possible threat to the validity of the results of our experiments is that we cannot guarantee 100% accuracy in segmenting SE texts and recognizing POS taggers based on Stanford CoreNLP. However, existing work has reported that the accuracy of off-the-shelf NLP tools is acceptable when analyzing texts with the context of proper sentences and grammatical structures, instead of analyzing fragmented source code [44]. With additional preprocessing, we think the quality of our analyzed SE texts is able to hold usable sentence structures for our approach. During our observations, we found no obvious errors from the output of Stanford CoreNLP either. Another possible threat is that our observations are not thorough and complete enough to fully exploit how developers express their sentiments on SE texts, and thus our defined rules cannot cover all misjudged sentiments found by our observations on the evaluated datasets. Still, we think these rules defined in this paper make a good start because with their help, our approach is able to achieve an overall better performance compared to the baseline approaches. We plan to make a deeper and more comprehensive study guided by psychology and sociology theories by consulting existing work (e.g., mental workload assessment [34]) in future.

**External Threats.** Our work is based on four datasets containing 7,190 SE texts in total with manually labeled sentiments. The size of our experiment is not large, but we still consider our findings relevant because the four datasets come from two existing work [13, 36] and are generated from three different online tools for software development (Stack Overflow, App Reviews, and JIRA). The two datasets from Stack Overflow are able to represent developers' typical interactions through SE texts due to the wide popularity of Stack Overflow. The other two datasets, unlike the previous two, contain SE texts that are almost labeled as either positive or negative sentiments. Thus, these two datasets are very helpful to verify whether SESSION overemphasizes the neutral sentiments (the majority sentiments in the two Stack Overflow datasets) in SE texts. Our evaluation shows that the performance of SESSION hardly decreases on the App Reviews and JIRA datasets, where SentiStrength-SE and Senti4SD (the two SE-customized baseline approaches) suffer a visible loss in their performance.

## VII. RELATED WORK

In this section, we focus our discussion of related research on sentiment analysis in the software engineering domain.

### A. Sentiment Analysis Tools Applied to SE

A comprehensive set of out-of-the-box sentiment analysis tools developed and used to detect sentiments can be found elsewhere [8, 9, 10]. Among these tools, SentiStrength [11], NLTK [39] and StanfordNLP [37] are common-used in SE domain. However, these tools do not perform well when applied to SE texts [12, 36, 40, 41] largely due to being trained on non-technical texts. Hence, some studies were conducted to improve the situation by utilizing SE texts, such as SentiCR [15], SentiStrenght-SE [12], and Senti4SD [13]. SentiCR is a supervised tool trained using Gradient Boosting Tree (GBT) [17] that is especially designed for code review comments. It generates feature vectors by computing TF-IDF [16] (Term Frequency - Inverse Document Frequency) of bag-of-words extracted from the input text. SentiStrenght-SE is a dictionary-based tool developed from SentiStrength by extending inherent dictionary with SE terms, which is the first SE-specific sentiment analysis. Senti4SD [13] is trained on a gold standard of about 4K questions, answers, and comments from Stack Overflow. It leverages three kinds of features when conducting sentiment classification tasks, including dictionary-based features (i.e., the dictionary used by SentiStrength), keyword-based features (i.e., uni-grams and bi-grams extracted from large scale Stack Overflow posts), and semantic features (based on the word embeddings trained on Stack Overflow posts). Unlike Senti4SD leveraging keyword-based features in the corpus, we paid more attention to analyze characteristics of SE texts and created a set of heuristics leveraging sentence structure information (e.g., identifying subjunctive clauses or distinguishing the meaning of polysemous words) based on our close observations. Furthermore, our approach is dictionary-based which can be more generalized to various SE texts, while learning-based methods need a large scale of labeled data to train their classifiers [15, 25, 36].

Apart from the discussed sentiment analysis tools designed to detect sentiment polarities (i.e., positivity, negativity, and neutrality) of a given text, Islam et al. [14] proposed a dictionary-based tool that can detect excitement, stress, depression, and relaxation expressed in software engineering text. To better assess the sentiment scores, their approach also integrated with a set of heuristics for sensing arousal, but it does not explicitly take advantage of the sentence structures from SE texts, while in this paper we use these sentence structures as the basis of our filter-adjust rules for our approach.

### B. Sentiment Analysis Application in SE

In recent years, sentiment analysis is receiving increasing attention as part of human factors of SE [18] and has been widely applied in SE tasks [19-28]. A number of studies applied sentiment analysis in the collaborative online environment (e.g., GitHub, JIRA, Stack Overflow, and App store) presented as follows: Pletea et al. [29] mined emotions from security-related discussions around commits and pull request on GitHub, and found that more negative emotions are expressed in security-related discussions than in other discussions. Guzman et al. [30] used dictionary-based sentiment analysis to detect sentiment expressed in commit comments of six OSS projects in GitHub and showed that the projects with more distributed teams tend to have a higher positive polarity in their emotional content. Mantyla et al. [21] analyzed 700,000 JIRA issues containing 2,000,000 comments with VAD (Valence, Arousal, and Dominance) metrics. The result indicated that different type issues reports (e.g., Feature Request, Improvement, and Bug Report) have a fair variation of Valence, while an increase in issue priority (e.g., from Minor to Critical) typically increases Arousal. Ortu et al. [22] analyzed the relation between sentiments, emotions and politeness of developers in more than 560K JIRA comments with the time to fix a JIRA issue. They found that the happier developers (expressing emotions such as JOY and LOVE in their comments) tend to fix an issue in a shorter time. Calefato et al. [32] quantitatively analyze emotions of a set of over 87K questions from the Stack Overflow finding that successful questions usually adopt a neutral emotional style. Canfora et al. [34] showed that users feedback contains usage scenarios, bug reports, and feature requests, that can help app developers to accomplish software maintenance and evolution tasks.

However, we also need to point out that the precision and reliability of the current sentiment analysis tools in SE domain are still less than satisfaction [20, 36]. One possible reason is that many prior works leverage off-the-shelf sentiment analysis tools (such as SentiStrength [11]) built on texts that are irrelevant to the SE domain, while proposing an SE-specified sentiment analysis is challenging [36]. Therefore, in this paper we choose to first exploit the uniqueness of sentiment expressions in the SE text, and then propose our approach by integrating our filter-adjust rules into SentiStrength.

## VIII. CONCLUSIONS AND FUTURE WORK

A growing body of work applies sentiment analysis on SE texts to enhance software development and program comprehension. However, current automated sentiment analysis, even including two SE-customized approaches, cannot provide reliable results on SE texts. Thus, we first observed and found that the expression of sentiments in SE texts are more indirect and dispersed compared to texts from common social network. We then proposed a set of filter and adjust rules based on sentence structures inside SE texts, and combine these heuristics with the mainstream dictionary-based approach called SentiStrength. Our evaluation based on four different datasets showed that our approach has the overall better performance and generalizability than three baseline approaches. Our tool is now publicly available [43].

The possible directions of our future work are as follows: (1) we plan to further explore how and why developers express their sentiments in SE texts under the guidance of related psychology and sociology theories so that we can fine-tune and enrich our filter-adjust rules accordingly; (2) we plan to further improve sentiment analysis on SE texts by proposing an SE-Specified dictionary by consulting existing work (e.g., [45]); (3) we plan to further explore whether the expressed sentiments on SE texts, if correctly identified, explicitly correlate to the status of ongoing software development from multiple perspectives.

ACKNOWLEDGMENT

This work is jointly supported by the National Key Research and Development Program of China (No. 2019YFE0105500) and the Research Council of Norway (No. 309494), as well as the National Natural Science Foundation of China (Grants No.62072227, 61802173, and 61690204), Intergovernmental Bilateral Innovation Project of Jiangsu Province (BZ2020017), and the Collaborative Innovation Center of Novel Software Technology and Industrialization.